\documentclass[%
 aip,
 sd,%
 amsmath,amssymb,
 reprint,%
]{revtex4-1}

\usepackage{graphicx}
\usepackage{dcolumn}
\usepackage{bm}

\usepackage{docs}%
\usepackage[colorlinks=true,linkcolor=blue]{hyperref}%
\usepackage{amssymb}

\newcommand\beq{\begin{equation}}
\newcommand\eeq{\end{equation}}

\newcommand{\xv}{{\hat {\bf x}}}
\newcommand{\yv}{{\hat {\bf y}}}

\begin{document}

\title{Optical isolation via unidirectional resonant photon tunneling}%

\author{Massimo Moccia}
\affiliation{Waves Group, Department of Engineering, University of Sannio, I-82100 Benevento, Italy}

\author{Giuseppe Castaldi}
\affiliation{Waves Group, Department of Engineering, University of Sannio, I-82100 Benevento, Italy}

\author{Vincenzo Galdi}
\email{vgaldi@unisannio.it} 
\affiliation{Waves Group, Department of Engineering, University of Sannio, I-82100 Benevento, Italy}

\author{Andrea Al\`u}
\affiliation{Department of Electrical and Computer Engineering, The University of Texas at Austin, Austin, TX 78712, USA}

\author{Nader Engheta}
\affiliation{Department of Electrical and Systems Engineering, University of Pennsylvania, Philadelphia, PA 19104, USA}

\date{\today}

\begin{abstract}
We show that tri-layer structures combining epsilon-negative and magneto-optical material layers can exhibit unidirectional resonant photon tunneling phenomena that can discriminate between circularly-polarized (CP) waves of given handedness impinging from opposite directions, or between CP waves with different handedness impinging from the same direction. This physical principle, which can also be interpreted in terms of a Fabry-Perot-type resonance, may be utilized to design compact optical isolators for CP waves. Within this framework, we derive simple analytical conditions and design formulae, and quantitatively assess the isolation performance, also taking into account the unavoidable imperfections and nonidealities.
\end{abstract}

\maketitle


\section{Introduction}

Currently, there is a growing interest in engineering {\em nonreciprocal} light propagation effects, \cite{Potton:2004} in order to design 
devices such as isolators and circulators that can prevent undesired backward reflection and interference effects in optical circuits.
 
A typical way to attain such effects relies on {\em magneto-optical} (MO) materials \cite{Kotov:2010} which, in the presence of a (static) bias magnetic field that breaks the time-reversal symmetry, are characterized by different propagation constants for
circularly-polarized (CP) waves with right- and left-handedness (RCP and LCP, respectively). \cite{Landau:1984} However, the MO activity of candidate materials, such as bismuth iron garnet (BIG), \cite{Adachi:2000,Tepper:2003} is quite weak at optical frequencies, thereby leading to rather bulky implementations. Therefore, several strategies have been explored in order to achieve significant response enhancements, all essentially based on the coupling of MO effects with resonant phenomena. 

For instance, one-dimensional magneto-photonic crystals combining dielectric and MO material layers have been proposed, in which Bragg scattering enables enhanced Faraday and Kerr rotation effects \cite{Inoue:1998if,Sakaguchi:1999kj,Steel:2000di,Steel:2000iy,Li:2005iq} and one-way total reflection. \cite{Yu:2007ir} Moreover, optical isolation effects have been demonstrated in MO photonic crystal slabs in the form of perforated MO films supporting guided resonances. \cite{Fang:2011cd} Especially promising is the MO-based hybridization of plasmonic nanostructures, in the form of structured films supporting surface plasmon polaritons \cite{Belotelov:2007,Khanikaev:2007dh,Wurtz:2008,Ctistis:2009,Belotelov:2009ci,Zhu:2011jw} or surface spoof plasmons, \cite{Khanikaev:2010de} as well as nanoparticles \cite{Jain:2009,Steinberg1,Wang:2011,Steinberg2,Steinberg3} and nanowires \cite{Chin:2013bg} supporting localized surface plasmon resonances.
Also, resonant tunneling effects have been demonstrated in multilayered heterostructures containing metallic and MO constituents, \cite{Chen:2012et} as well as dielectric and MO metals. \cite{Dong:2010ez,Moccia:2014}
In the same context, of special interest are also some theoretical studies on multilayered structures containing magnetically-biased constituents aimed at achieving unidirectional response \cite{Figotin:2003kg} and minimizing the detrimental effects of losses. \cite{Figotin:2008hr}

Alternatively, also worth of mention are strategies that do not rely on an external magnetic-field bias, but instead on 
nonlinear media, \cite{Scalora:1994cq,Tocci:1995,Gallo:2001,Miroshnichenko:2010} or time-dependent modulation of the refractive index. \cite{Yu:2009fr}

More recently, the use of metamaterials has been proposed for the design of compact optical isolators. In particular, a multilayer based on nonlinear-plasma and double-negative metamaterial constituents was considered in Ref. \onlinecite{Kong:2013cm}. On the other hand,  the concept of MO-active {\em epsilon-near-zero} (ENZ) metamaterial was put forward in Ref. \onlinecite{Davoyan:2013dq} in order to achieve optical isolation of CP waves, and two possible designs were explored, based on metal-MO multilayers and near-cutoff waveguides filled by an MO material.  

Inspired by the idea in Ref. \onlinecite{Davoyan:2013dq}, we explore here a different physical route to optical isolation, based on unidirectional resonant photon tunneling through simple tri-layers containing metallic and MO constituents. Our idea essentially relies on the large body of results available on resonant photon tunneling phenomena in multilayered heterostructures combining {\em single-negative} (SNG, i.e., with negative permittivity or permeability) and {\em double-positive} (DPS, i.e., with positive permittivity and permeability) material constituents. \cite{Zhou:2005dr,Feng:2009jk,Castaldi:2011kg,Castaldi:2011ka,Cojocaru:2011tl} By substituting in these heterostructures a DPS material with an MO material, we render the resonance conditions sensitive to the CP handedness, thereby creating an adjustable contrast between the transmission responses of forward- and backward-propagating CP waves of a given handedness.

Accordingly, the rest of the paper is laid out as follows. In Sect. \ref{Sect:Main}, we introduce the problem geometry and formulation, and outline the main analytical derivations, with specific reference to a configuration featuring an MO layer sandwiched between two epsilon-negative (ENG) material layers. In Sect. \ref{Sect:Results}, we discuss some representative numerical results, in order to illustrate and
quantitatively assess the isolation properties, also taking into account the unavoidable fabrication tolerances, as well as material dispersion and loss effects. In Sect. \ref{Sect:Alternative}, we briefly discuss possible alternative configurations. Finally, in Sect. \ref{Sect:Conclusions}, we provide some concluding remarks and hints for future research.

\section{Unidirectional resonant photon tunneling in ENG-MO-ENG tri-layers}
\label{Sect:Main}

\subsection{Problem geometry and formulation}

%
\begin{figure}
\begin{center}
\includegraphics [width=8.5cm]{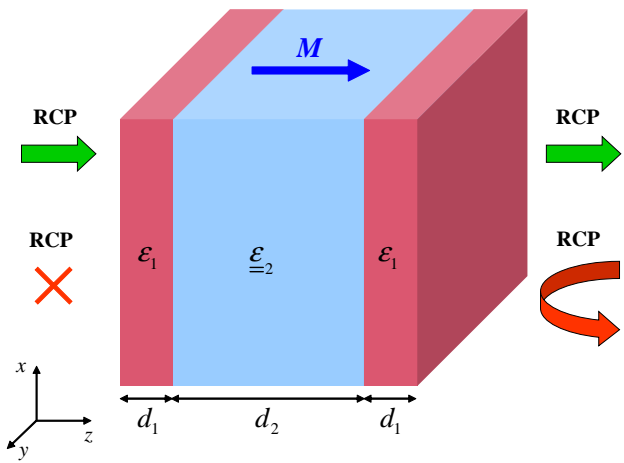}
\end{center}
\caption{(Color online) Problem schematic: A tri-layer structure of infinite extent in the transverse $(x,y)$ plane, consisting of a slab of MO material of thickness $d_2$ and relative permittivity tensor ${\underline {\underline \varepsilon}}_2$ [assuming a static bias magnetization along the $z$-direction, cf. (\ref{eq:eps2})] sandwiched between two identical ENG layers of thickness $d_1$ and relative permittivity $\varepsilon_1$ [with $\mbox{Re}(\varepsilon_1)<0$], immersed in vacuum. In the ideal lossless case, at selected resonant frequencies, an RCP illumination is totally transmitted (thick green arrows) when impinging from the left, and strongly reflected when impinging from the right (red thick arrow). For the LCP case, the directions are reversed.}
\label{Figure1}
\end{figure}

In a series of papers, \cite{Zhou:2005dr,Feng:2009jk,Castaldi:2011kg,Castaldi:2011ka,Cojocaru:2011tl} it was shown that multilayered heterostructures containing SNG material constituents suitably combined with DPS materials, in the ideal lossless case, can become {\em perfectly transparent} at certain frequencies. Under these resonant conditions, the energy can tunnel through the structure despite the {\em inherent opaqueness} of the SNG material layers. As previously mentioned, our basic idea is to render these resonance conditions sensitive to the CP handedness by replacing a DPS material constituent with an MO one.
With a view towards applications to optical frequencies, here we do not consider configurations containing {\em mu-negative} (MNG) materials (such as the one in Ref. \onlinecite{Feng:2009jk}), and focus instead on those containing ENG materials, \cite{Zhou:2005dr,Castaldi:2011kg,Cojocaru:2011tl} which are naturally available  at these frequencies, e.g., in the form of noble metals.

In particular, we begin considering the tri-layer geometry in Fig. \ref{Figure1}, which features an MO layer of thickness $d_2$ sandwiched between two identical ENG layers of relative permittivity $\varepsilon_1$ [with $\mbox{Re}(\varepsilon_1)<0$] and thickness $d_1$, immersed in vacuum. All materials are assumed to be nonmagnetic ($\mu=1$), the structure is of infinite extent in the transverse ($x,y$) plane, and a $z-$directed static bias magnetization is considered. Accordingly, in the time-harmonic [$\exp(-i\omega t)$] regime, the MO material is described by a {\em gyrotropic} relative permittivity tensor \cite{Landau:1984}
\beq
{\underline {\underline \varepsilon}}_2=\left[
\begin{array}{ccc}
\varepsilon_{MO} & -i\alpha & 0\\
i\alpha & \varepsilon_{MO} & 0\\
0 & 0 & \varepsilon_{\perp}
\end{array}
\right].
\label{eq:eps2}
\eeq
The above configuration differs from that studied (among others) in Ref. \onlinecite{Cojocaru:2011tl} by the presence of the MO layer in place of an isotropic DPS layer. In what follows, we analytically study the electromagnetic response of this ENG-MO-ENG tri-layer to CP waves propagating along the magnetization direction $z$, i.e., in the so-called Faraday configuration. In particular, we determine the resonant tunneling conditions under which an incident electromagnetic field with CP of given handedness (RCP, in Fig. \ref{Figure1}) is totally transmitted when impinging from one direction (left, in Fig. \ref{Figure1}), and strongly reflected when impinging from the opposite direction (right, in Fig. \ref{Figure1}). This effect can be exploited in order to design compact optical isolators for CP waves.

Incidentally, a similar configuration was also recently studied in Ref. \onlinecite{Chin:2013bg} for linearly-polarized excitation. In particular, via fully-numerical studies, it was shown that, at certain special frequencies, an incident linearly-polarized field could give rise to purely CP transmitted or reflected waves. Our analytical results below also provide further insights into this phenomenon. 

\subsection{General solution and conditions for unidirectional tunneling}
Referring to Sect. \ref{Sec:Losses} for more realistic scenarios, we start considering the ideal lossless case, where the relevant constitutive parameters $\varepsilon_1$, $\varepsilon_{MO}$ and $\alpha$ are all real-valued, and we calculate the electromagnetic response for an impinging RCP excitation with electric field
\beq
{\bf E}^{(i,L)}_{RCP}\left(z\right)=\exp\left(i k z\right)\left(\xv+i\yv\right),
\label{eq:EiL}
\eeq
propagating in the forward $z$ direction (i.e., normally impinging from the left, in Fig. \ref{Figure1}). In (\ref{eq:EiL}), and henceforth,
$k=\omega/c=2\pi/\lambda$ denotes the vacuum wavenumber (with $c$ and $\lambda$ being the corresponding wavespeed and wavelength, respectively), boldface symbols identify vector quantities, and $\xv$ and $\yv$ are the $x-$ and $y-$directed unit vectors, respectively.

It is well-known that, in the assumed Faraday configuration, the fields inside the MO layer can be represented in terms of forward- and backward-propagating CP eigenwaves with electric fields \cite{Landau:1984}
\begin{subequations}
\begin{eqnarray}
{\bf E}_{RCP}^{(FW)}(z)&=&\exp\left(i \beta_{(+)} z\right)\left(\xv+i\yv\right),\\
{\bf E}_{RCP}^{(BW)}(z)&=&\exp\left(-i \beta_{(-)} z\right)\left(\xv-i\yv\right),\\
{\bf E}_{LCP}^{(FW)}(z)&=&\exp\left(i \beta_{(-)} z\right)\left(\xv-i\yv\right),\\
{\bf E}_{LCP}^{(BW)}(z)&=&\exp\left(-i \beta_{(+)} z\right)\left(\xv+i\yv\right),
\end{eqnarray}
\end{subequations}
and different propagation constants
\beq
\beta_{(+)}=k\sqrt{\varepsilon_{MO}+ \alpha},~~
\beta_{(-)}=k\sqrt{\varepsilon_{MO}- \alpha}.
\eeq

For the assumed forward-propagating RCP excitation, the electric field in the various regions of the structure in Fig. \ref{Figure1} can be written in terms of linear combinations of forward-propagating RCP waves and backward-propagating LCP waves, which become evanescent in the ENG layers, viz.,
\beq
{\bf E}\left(z\right)=e\left(z\right) \left(\xv+i\yv\right),
\label{eq:EE}
\eeq
with
\begin{subequations}
\begin{eqnarray}
e\left(z\right)&=& \exp\left(i k z\right)+B_0\exp\left(-i k z\right),~~z<0,\\
e\left(z\right)&\!=\!& A_1 \exp\left(-\gamma z\right) \!+\!B_1\exp\left(\gamma z\right)\!,~ 0<z<d_1,\\
e\left(z\right)&=& A_2 \exp\left(i \beta_{(+)} z\right) +B_2\exp\left(-i \beta_{(+)} z\right),\nonumber\\
&& d_1<z<d_1+d_2,\\
e\left(z\right)&=&A_3 \exp\left(-\gamma z\right) +B_3\exp\left(\gamma z\right),\nonumber\\
&& d_1+d_2<z<2d_1+d_2,\\
e\left(z\right)&\!=\!&A_4 \exp\left[ik \left(z-2d_1-d_2\right)\right]\!,~z>2d_1+d_2,
\end{eqnarray}
\label{eq:ee}
\end{subequations}
where $\gamma=k\sqrt{-\varepsilon_1}$ indicates the attenuation constant of the ENG material. The eight unknown coefficients $A_1$, $A_2$, $A_3$, $A_4$, $B_0$, $B_1$, $B_2$, and $B_3$ can be calculated by enforcing the  of electric and magnetic tangential fields at the four interfaces $z=0$, $z=d_1$, $z=d_1+d_2$, and $z=2d_1+d_2$, with the magnetic field derived from  (\ref{eq:EE}) and (\ref{eq:ee}) via the relevant Maxwell's curl equation
\beq
{\bf H}\left(z\right)=\frac{1}{i\omega\mu_0}\nabla \times {\bf E}\left(z\right),
\eeq
where $\mu_0$ denotes the vacuum permeability. The arising $8\times 8$ linear system of equations can be solved analytically in a cumbersome but straightforward fashion. We focus, in particular, on the coefficients $B_0$ and $A_4$, which play the role of reflection and transmission coefficients, respectively,
\begin{subequations}
\beq
R_L=B_0=\frac{N_{RL}}{D_L},
\eeq
\begin{eqnarray}
N_{RL}&=&2 \tau  \gamma \beta_{(+)}  \left(k^2+\gamma ^2\right)\nonumber\\ 
&\!-\!&\xi_{(+)}
\left[\beta _{(+)}^2
\left(k^2 \tau ^2+\gamma ^2\right) 
\!-\!\gamma^2 \left(k^2\!+\! \gamma ^2\tau ^2\right)\right]\!,\\
D_L&=&
2 \gamma \beta _{(+)} \left(k \tau +i \gamma\right) \left(k+i \tau  \gamma \right)\nonumber\\ 
&-&
\xi_{(+)}\left[\beta_{(+)}^2\left(k \tau +i \gamma\right)^2+\gamma ^2 \left(\tau  \gamma -i k \right)^2 \right],
\label{eq:DL}
\end{eqnarray}
\label{eq:RL}
\end{subequations}
\beq
T_L=A_4=\frac{2 i k \beta_{(+)} \gamma ^2   \left(\tau ^2-1\right) \sec\left(\beta_{(+)} d_2\right)}{D_L},
\label{eq:TL}
\eeq
with the subscript $_L$ denoting the incidence from the left, and
\beq
\tau=\tanh\left(\gamma d_1\right),~~~\xi_{(+)}=\tan\left(\beta_{(+)}d_2\right).
\eeq
From (\ref{eq:RL}), we can readily derive the resonant photon tunneling conditions by zeroing the numerator $N_{RL}$ of the reflection coefficient. For given values of frequency, ENG parameters $\varepsilon_1$ and $d_1$ and MO constitutive parameters $\varepsilon_{MO}$ and $\alpha$, this always yields a {\em discrete infinity} of solutions,
\begin{eqnarray}
d_2&=&\frac{1}{\beta_{(+)}}\nonumber\\
&\times&\left\{
\arctan
\left[
\frac{2 \tau \beta _{(+)} \gamma  \left(k^2+\gamma ^2\right)}{\beta _{(+)}^2\left(k^2 \tau ^2+\gamma ^2\right)-\gamma ^2 \left(k^2+ \gamma ^2\tau^2\right)}
\right]\right.\nonumber\\
&+&n
\Bigl.\pi\Biggr\},~~~n\in \mathbb{N},
\label{eq:d2n}
\end{eqnarray}
which correspond to different electrical thicknesses of the MO layer. It can be verified that the solutions in (\ref{eq:d2n}) are not roots of the denominator $D_L$ in (\ref{eq:DL}), and hence they are actual zeros of the reflection coefficient for RCP incidence from the left. It can also be verified that the expression in (\ref{eq:d2n}) can be recast in a form analogous to that obtained in Ref. \onlinecite{Cojocaru:2011tl} for the isotropic case.

Assuming now a backward-propagating RCP illumination (i.e., normally impinging from the right, in Fig. \ref{Figure1}), with electric field
\beq
{\bf E}^{(i,R)}_{RCP}\left(z\right)=\exp\left(-i k z\right)\left(\xv-i\yv\right),
\eeq
we can repeat the procedure above with only slight changes. Omitting the details for brevity, we limit ourselves to report the final results for the reflection and transmission coefficients,
\begin{subequations}
\beq
R_R=\frac{N_{RR}}{D_R},
\eeq
\begin{eqnarray}
N_{RR}&=&2 \tau  \gamma \beta_{(-)}  \left(k^2+\gamma ^2\right) \nonumber\\
&\!-\!\!&\xi_{(-)}
\!\left[\beta_{(-)}^2
\!\left(k^2 \tau ^2\!+\!\gamma ^2\right) 
\!-\!\gamma^2 \left(k^2\!+\! \gamma ^2\tau ^2\right)\!\right]\!,\\
D_R&=&
2 \gamma \beta _{(-)} \left(k \tau +i \gamma\right) \left(k+i \tau  \gamma \right)\nonumber\\ &\!-\!&
\xi_{(-)}\left[\beta_{(-)}^2\left(k \tau \!+\!\ \gamma\right)^2+\gamma ^2 \left(\tau  \gamma -i k \right)^2 \right]\!,
\end{eqnarray}
\label{eq:RR}
\end{subequations}
\beq
T_R=\frac{2 i k \beta_{(-)} \gamma ^2   \left(\tau ^2-1\right) \sec\left(\beta_{(-)} d_2\right)}{D_R},
\label{eq:TR}
\eeq
where the subscript $_R$ denotes the incidence from the right, and
\beq
\xi_{(-)}=\tan\left(\beta_{(-)} d_2\right).
\eeq
By simple inspection, it can readily be verified that the expressions in (\ref{eq:RR}) and (\ref{eq:TR}) differ from their counterparts in (\ref{eq:RL}) and (\ref{eq:TL}) (pertaining to the incidence from the left) only by the presence of $\beta_{(-)}$ in place of $\beta_{(+)}$, i.e., as an effect of the MO activity. As previously mentioned, at optical frequencies, the MO activity of typical available materials (e.g., BIG \cite{Adachi:2000,Tepper:2003}) is rather weak, resulting in $\alpha\ll \varepsilon_{MO}$. This implies a quite small difference in the two propagation constants,
\beq
\Delta \beta\equiv \beta_{(+)}-\beta_{(-)}\approx \frac{\alpha k}{\sqrt{\varepsilon_{MO}}}\ll k,
\eeq
so that large interaction volumes are usually required in order to achieve sensible differences in the phase accumulation.
In what follows, we show that, in our scenario, the MO activity can be greatly enhanced by the underlying resonant phenomenon, so that significant differences in the transmission response of a given CP handedness for opposite incidence directions (or, equivalently, of two CP waves with different handedness impinging from the same direction) may be attained with compact (wavelength-sized) structures.

\section{Representative numerical results}
\label{Sect:Results}

%
\begin{figure}
\begin{center}
\includegraphics [width=8.5cm]{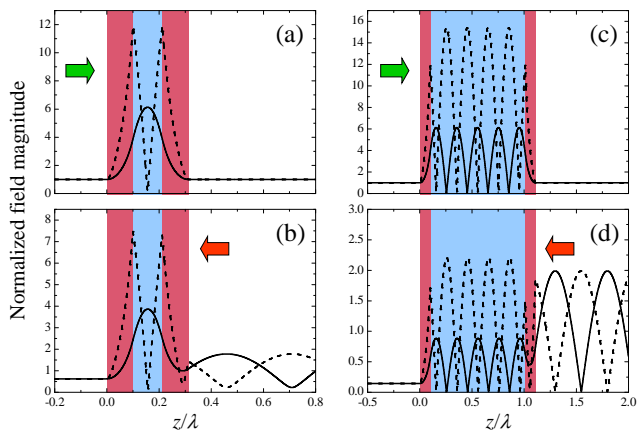}
\end{center}
\caption{(Color online) (a), (b) Resonant electric (solid) and magnetic (dashed) field-magnitude distributions along the $z$ axis (normalized by the incident fields) for a configuration as in Fig. \ref{Figure1}, with $\varepsilon_1=-10$, $d_1=0.1 \lambda$, $\varepsilon_{MO}=6.25$, $\alpha=0.06$, $d_2=0.112\lambda$ [i.e., $n=1$ in (\ref{eq:d2n})], for normally-incident RCP illumination from the left and right, respectively (as also indicated by the thick arrows).
(c), (d) Same as (a), (b), but for $d_2=0.908\lambda$ [i.e., $n=5$ in (\ref{eq:d2n})]. Different colors (consistent with Fig. \ref{Figure1}) are used in order to visually delimit the various material layers. Also, please note the difference in scale of the different panels.}
\label{Figure2}
\end{figure}

\subsection{Unidirectional tunneling}
In order to qualitatively illustrate and understand the unidirectional resonant photon tunneling phenomenon, we show in Fig. \ref{Figure2} the electric and magnetic field distributions pertaining to two representative structures, assuming an RCP excitation normally impinging from the left and right. In these examples, the ENG parameters ($\varepsilon_1=-10$ and $d_1=0.1\lambda$) were merely selected 
for the sake of simplicity of visualization, whereas the MO constitutive parameters ($\varepsilon_{MO}=6.25$ and $\alpha=0.06$) represent realistic values for BIG at the telecom wavelength $\lambda=1550$nm, \cite{Adachi:2000,Tepper:2003} also utilized in other related studies. \cite{Fang:2011cd,Davoyan:2013dq}
The MO layer electrical thickness $d_2/\lambda$ was chosen according to (\ref{eq:d2n}), i.e., in order to achieve total transmission for incidence from the left. 

More specifically, Fig. \ref{Figure2}(a) shows the resonant field distributions along the $z$ axis for incidence from the left, assuming a thickness $d_2=0.112\lambda$, i.e., considering the smallest positive solution ($n=1$) in (\ref{eq:d2n}). Total transmission is clearly visible, with evanescent field growth in the left ENG layer, and the 
electric and magnetic fields peaked at the center of the MO layer and at its boundaries with the ENG layers, respectively. We point out that, although in the topical literature \cite{Cojocaru:2011tl,Chen:2012et} such resonant phenomena are referred to as ``tunneling,''  the response resembles that of a Fabry-Perot resonant cavity, effectively formed by the highly reflective properties of the ENG end layers.
For the same structure, Fig. \ref{Figure2}(b) shows the response in the case of incidence from the right, from which a sensible (albeit not extreme) difference can be observed. A standing-wave pattern is now clearly visible in the incidence region, indicating a moderate reflection, and the transmission is accordingly not full ($|T_R|^2=0.394$).

Figures \ref{Figure2}(c) and \ref{Figure2}(d) show the same responses for a moderately larger structure ($d_2=0.908\lambda$) operating at a higher-order resonance [$n=5$ in (\ref{eq:d2n})]. In addition to the understandably more complicated standing-wave patterns in the MO layer, a much more marked difference between the output responses can be observed. In particular, while the incidence from the left still yields total transmission [Fig. \ref{Figure2}(c)], a significant reflection is now experienced for incidence from the right [Fig. \ref{Figure2}(d)], resulting in a quite low transmission ($|T_R|^2=0.021$; please note the difference in scale of the different panels). 

%
\begin{figure}
\begin{center}
\includegraphics [width=8.5cm]{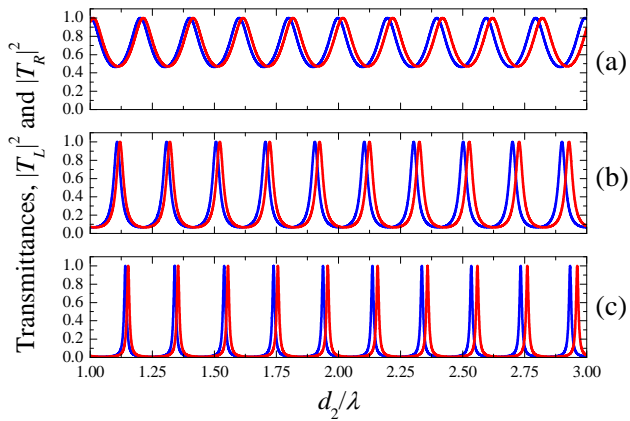}
\end{center}
\caption{(Color online) Transmittances for RCP illumination normally-incident from the left [$|T_L|^2$, cf. (\ref{eq:TL}); blue curves] and right [$|T_R|^2$, cf. (\ref{eq:TR}); red curves] as a function of the MO layer electrical thickness $d_2/\lambda$, for a structure as in Fig. \ref{Figure2} but with $d_1=0.01 \lambda$ and various values of $\varepsilon_1$. (a), (b), (c) $\varepsilon_1=-1$, $\varepsilon_1=-50$, and $\varepsilon_1=-100$, respectively.}
\label{Figure3}
\end{figure}

\subsection{Isolation properties}
\label{eq:Isolation}
For a more quantitative and thorough illustration and characterization of the phenomenon, Fig. \ref{Figure3} compares the transmittances for incidence from the left and right ($|T_L|^2$ and $|T_R|^2$) as a function of the electrical thickness of the MO layer $d_2/\lambda$, for a given ENG layer thickness ($d_1=0.01\lambda$) and three representative values of the relative permittivity $\varepsilon_1$.
In all three cases, we observe a series of equispaced total-transmission peaks alternated with transmission minima, with a shift between responses for opposite incidence directions (blue and red curves), which progressively increases with the MO layer thickness. However, for low (absolute) values of $\varepsilon_1$ [i.e., mildly reflective ENG layers, cf. Fig. \ref{Figure3}(a)], the peaks are rather wide and the dynamic range is only moderate, resulting in less than optimal isolation properties. For increasing (absolute) values of $\varepsilon_1$ [cf. Figs. \ref{Figure3}(b) and \ref{Figure3}(c)], the peaks tend to become more selective and to exhibit larger dynamic ranges, encompassing very low transmission levels. For instance, assuming $\varepsilon_1=-100$ (a value consistent with the real-part of relative permittivity exhibited by noble metals at optical frequencies \cite{Johnson:1972}), we observe from Fig. \ref{Figure3}(c) that already for thickness values $\sim 2\lambda$ the peak selectivity and shift are sufficient to ensure that total transmission for incidence from a given side is accompanied by very low ($<0.1$) transmittance for incidence from the opposite side.  

%
\begin{figure}
\begin{center}
\includegraphics [width=8.5cm]{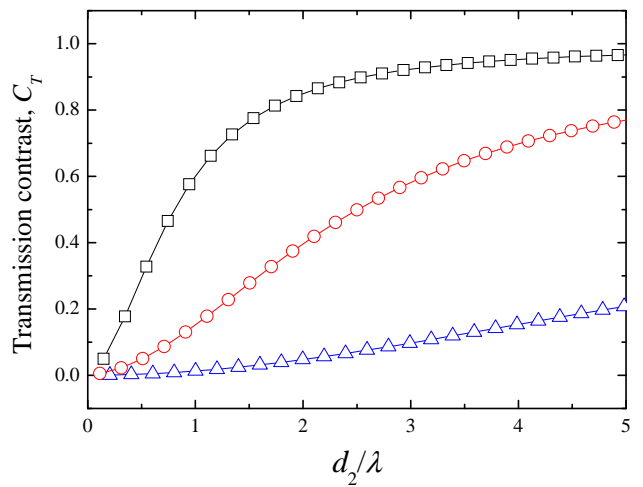}
\end{center}
\caption{(Color online) As in Fig. \ref{Figure3}, but transmission contrast [cf. (\ref{eq:CT})] sampled at values of $d_2/\lambda$ that yield total transmission for incidence from the left (i.e., $|T_L|=1$). Square, circle, and triangle markers pertain to $\varepsilon_1=-100$, $\varepsilon_1=-50$, and $\varepsilon_1=-1$, respectively. Connecting lines are guides to the eye only.}
\label{Figure4}
\end{figure}

For a more quantitative assessment of the isolation efficiency, following Refs. \onlinecite{Gallo:2001,Kong:2013cm}, we consider the {\em transmission contrast}
\beq
C_T=\frac{\left|T_L\right|^2-\left|T_R\right|^2}{\left|T_L\right|^2+\left|T_R\right|^2},
\label{eq:CT}
\eeq
which parameterizes the capability to discriminate between the two incidence directions. Ideally, in our configuration, we would like to achieve $|T_L|=1$ and $T_R=0$, i.e., $C_T=1$. For the same configuration and parameters as in Fig. \ref{Figure3}, we show in Fig. \ref{Figure4} the transmission contrast sampled at values of $d_2/\lambda$ that yield total transmission for incidence from the left (i.e., $|T_L|=1$). It can be observed that the transmission contrast generally increases with the electrical thickness of the MO layer, and the growth rate is more pronounced for larger (absolute) values of $\varepsilon_1$. Overall, for $\varepsilon_1=-100$, satisfactorily high transmission-contrast values $C_T\gtrsim 0.8$ may be obtained with reasonably compact structures of thickness $\sim 2\lambda$.

%
\begin{figure}
\begin{center}
\includegraphics [width=8.5cm]{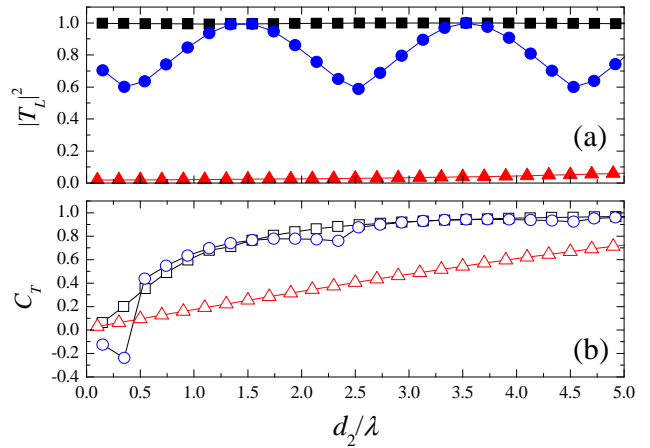}
\end{center}
\caption{(Color online) As in Fig. \ref{Figure4} for $\varepsilon_1=-100$, but considering both the transmittance for incidence from the left (a) and the transmission contrast (b), and assuming a round-off approximation of the nominal resonant values of $d_2/\lambda$ in (\ref{eq:d2n}). 
Square, circle, and triangle markers pertain to round-off approximation to three, two, and one decimal places, respectively.}
\label{Figure5}
\end{figure}

%
\begin{figure}
\begin{center}
\includegraphics [width=8.5cm]{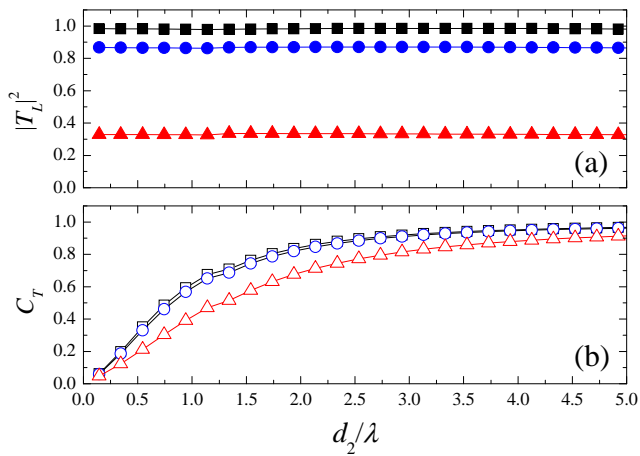}
\end{center}
\caption{(Color online) As in Fig. \ref{Figure5} (assuming a round-off approximation to three decimal places for $d_2/\lambda$), but considering a lossy ENG material. Square, circle, and triangle markers pertain to loss-tangent values of $0.001$, $0.01$, and $0.1$, respectively.}
\label{Figure6}
\end{figure}

\subsection{Effects of fabrication tolerances, dispersion and losses}
\label{Sec:Losses}
With a view toward practical realization aspects, given the resonant nature of the tunneling phenomenon, it is important to assess the performance sensitivity with respect to the unavoidable fabrication tolerances, as well as material dispersion and losses.

In connection with the first issue, the most critical parameter turns out to be the MO layer thickness $d_2$. In the presence of very selective peaks and minimal shifts in the $|T_L|^2$ and $|T_R|^2$ responses, small departures from the nominal resonant values in   (\ref{eq:d2n}) may severely curtail the isolation performance. As an illustrative example, with reference to the parameter configuration in Fig. \ref{Figure4} with $\varepsilon_1=-100$, Fig. \ref{Figure5} shows the detrimental effects due to the round-off approximation (with three to one decimal places) of the nominal resonant values of $d_2/\lambda$ in (\ref{eq:d2n}). This time, besides the transmission contrast, also the transmittance for incidence from the left is shown, since it generally differs from one. Basically, rounding-off to three decimal places turns out to ensure very similar results as the infinite-precision case in Fig. \ref{Figure4}. Worsening the round-off approximation to two decimal places mainly affects the transmittance response (with sensible deterioration up to $\sim 40\%$), with only mild effects in the transmission contrast. On the other hand, unacceptably low transmittance and contrast values are observed for a round-off approximation to one decimal place, indicating that the tunneling phenomenon essentially breaks down. From the above results, we can therefore expect a certain robustness of the isolation performance for fabrication tolerances on the order of $\sim 0.001\lambda$ to $\sim 0.01\lambda$, i.e., few nanometers at optical frequencies, which appear to be feasible within the current nanofabrication technologies.

%
\begin{figure}
\begin{center}
\includegraphics [width=8.5cm]{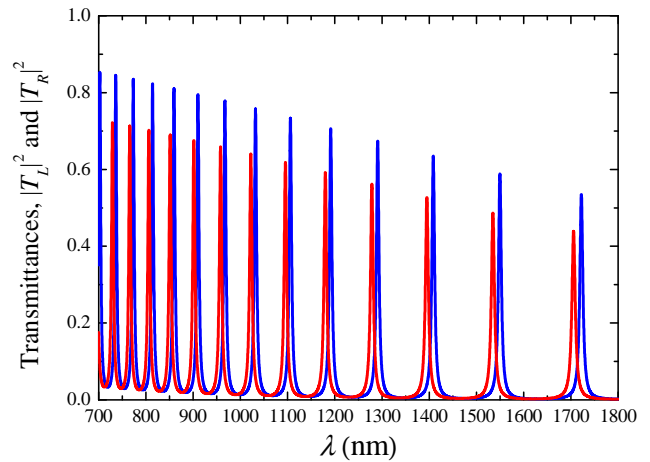}
\end{center}
\caption{As in Fig. \ref{Figure3}, but as a function of the wavelength. For the ENG material, the dispersive, lossy Drude model in (\ref{eq:Drude}) is assumed, with parameters $\omega_m=2\pi\times2.211$ PHz and $\nu=2\pi \times 5.133$ THz chosen so as to fit the experimental data available for silver. \cite{Johnson:1972} For the MO material, a nondispersive, lossy model is assumed, with parameters
 $\varepsilon_{MO} = 6.25 + i 0.0025$ and $\alpha=0.06-i0.0015$ consistent with other studies. \cite{Belotelov:2007} The layer thicknesses are chosen as $d_1=0.01\lambda_0$ and $d_2=1.945\lambda_0$ [i.e., $n=10$ in (\ref{eq:d2n}), neglecting losses], with $\lambda_0=1550$nm being the nominal design wavelength.}
\label{Figure7}
\end{figure}

%
\begin{figure}
\begin{center}
\includegraphics [width=8.5cm]{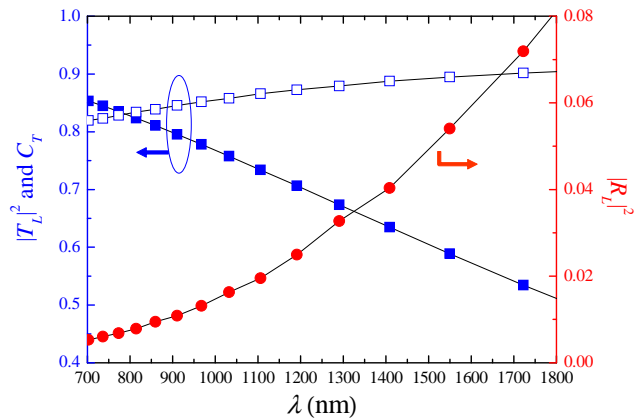}
\end{center}
\caption{(Color online) Parameters as in Fig. \ref{Figure7}. Transmittance for incidence from the left sampled at its peak values (full squares, left axis) and corresponding transmission contrast (empty squares, left axis). Also shown (circles, right axis) is the corresponding reflectance $|R_L|^2$. Connecting lines are guides to the eye only.}
\label{Figure8}
\end{figure}

Next, we move on to assessing the effect of losses. For the same parameters as in Fig. \ref{Figure5} (assuming a round-off approximation to three decimal places for $d_2/\lambda$), Fig. \ref{Figure6} shows the performance deterioration due to the inclusion of losses in the ENG layers, parameterized in terms of the loss-tangent. It can be observed that the performance is quite robust up to loss-tangent values $\sim 0.01$, whereas the transmittance significantly drops down for loss-tangent values $\sim 0.1$.
For a more realistic example, we show in Fig. \ref{Figure7} the transmittance spectra $|T_L|^2$ and $|T_R|^2$ in the near/short-infrared region obtained by assuming for the ENG material a Drude-type dispersive model, 
\beq
\varepsilon_1\left(\omega\right)=1-\frac{\omega_m^2}{\omega\left(\omega+i\nu\right)},
\label{eq:Drude}
\eeq
with parameters (given in the caption)  chosen so as to fit the experimental data available for silver. \cite{Johnson:1972} For the MO layer, we consider instead a nondispersive model with realistic loss parameters (given in the caption) consistent with other studies. \cite{Belotelov:2007} A series of narrow spectral peaks is observed, with amplidude decreasing with the wavelength, and a shift between the two spectra. In particular, at the telecom wavelength $\lambda=1550$nm (at which the structure was tuned), we obtain $|T_L|^2=0.588$ and $|T_R|^2=0.056$, with an overall thickness of $1.965\lambda\approx 3\mu$m. Even better performance can be obtained in the mid-infrared range, thanks to the lower losses exhibited by silver in that region. This is better quantified in Fig. \ref{Figure8}, which shows the transmittance for incidence from the left $|T_L|^2$ sampled at its peak values, and the corresponding transmission contrast and reflectance, as a function of the wavelength. It can be observed that the peak transmittance decreases almost linearly with the wavelength (with a corresponding increase of the reflectance), whereas the transmission contrast is uniformly closer to the ideal value ($C_T=1$), being always larger than $0.8$ and approaching $0.9$ at the telecom wavelength $\lambda=1550$nm. Within this spectral range, the reflectance is always below $0.08$. At wavelengths around $\lambda=800$nm, both the transmittance and the contrast are above $0.8$, and the reflectance is only $\sim 0.01$. As expected, in view of the underlying resonant phenomenon, the response is inherently narrow-bandwidth. However, this is a typical limitation of devices based on materials with weak MO-activity (such as BIG), whereby resonant phenomena are needed in order to enhance the nonreciprocal response.

We stress that our approach relies on the relatively low-loss character of the MO material (BIG) considered. In the presence of more significant losses in the MO material, the arising attenuation effects would need to be explicitly taken into account at the design stage, since they may be markedly sensitive to the CP handedness. 

\subsection{Some remarks}
\label{Sec:Remarks}
Overall, our results above indicate that the proposed unidirectional resonance mechanism is fairly robust, and yields reasonably high values of transmittance and contrast (with low reflectance) in wavelength-sized structures.

Although we have insofar dealt with the RCP case only, it is straightforward to show that formally analogous results hold for the LCP case too, but with a reversal of the incidence directions. In other words, for the same structures as above, an LCP illumination will be transmitted when impinging from the right, and reflected when impinging from the left. This implies that our proposed structure can separate two RCP and LCP waves impinging together along the same direction (by transmitting one and reflecting the other), and also explains the anomalous transmission/reflection effects observed in Ref. \onlinecite{Chen:2012et} for linearly-polarized illumination of a similar structure.

We emphasize that, though both relying on combinations of MO and metallic materials, the approach presented here and the one in Ref. \onlinecite{Davoyan:2013dq} are markedly different, as witnessed by the field distributions inside the structures (compare, e.g., Fig. \ref{Figure2} with Fig. 1d in Ref. \onlinecite{Davoyan:2013dq}). In our approach, for a given CP handedness, the forward and backward components are both propagating, and the isolation is achieved by judicious detuning of the corresponding tunneling resonances. Conversely, in Ref. \onlinecite{Davoyan:2013dq}, the backward component is evanescent, and the isolation is attained via resonant (Fabry-Perot-type) transmission of the forward propagating component. Moreover, from the structural viewpoint, our proposed configuration relies on a wavelength-sized MO-material layer sandwiched between two thin metal layers, which are arguably easier to fabricate than the multilayered structure (with deep-subwavelength layers of metallic and MO materials) proposed in Ref. \onlinecite{Davoyan:2013dq}.

%
\begin{figure}
\begin{center}
\includegraphics [width=8.5cm]{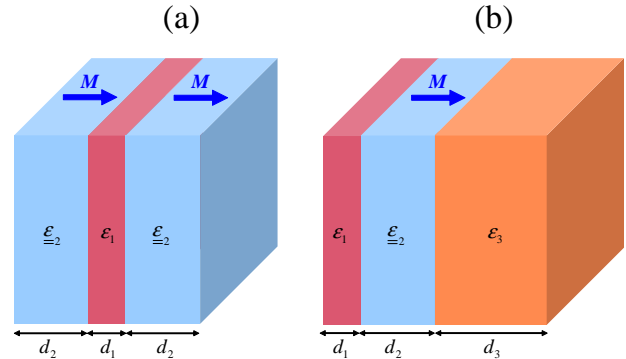}
\end{center}
\caption{(Color online) Possible alternative tri-layer configurations. (a) An ENG layer [$\mbox{Re}(\varepsilon_1)<0$] sandwiched between two identical MO material layers. (b) An MO material layer sandwiched between an ENG layer [$\mbox{Re}(\varepsilon_1)<0$] and a DPS layer [$\mbox{Re}(\varepsilon_3)>0$].}
\label{Figure9}
\end{figure}

\section{Possible alternative configurations}
\label{Sect:Alternative}
As previously mentioned, the tri-layer structure in Fig. \ref{Figure1} is only one of the possible candidate configurations. First, the ENG layers may be in principle replaced by highly-reflecting surfaces or metasurfaces.
Moreover, similar resonant effects can be attained in different configurations featuring at least three layers, one of which is ENG.
In what follows, we briefly discuss two possible alternatives.

We start considering the tri-layer structure in Fig \ref{Figure9}(a), featuring an ENG layer sandwiched between two identical MO layers. This structure, which is somehow ``complementary'' of the structure in Fig. \ref{Figure1}, is inspired by the study in Ref. \onlinecite{Zhou:2005dr} which demonstrated resonant photon tunneling in symmetrical DPS-ENG-DPS tri-layers. Also in this case, the substitution of the DPS layers with MO ones would render the phenomenon sensitive to the CP handedness. For brevity, the analytical derivations are not reported here, as they entail simple modifications of those in Ref. \onlinecite{Zhou:2005dr}. However, as an important difference with our results above, it is worth pointing out that now, for given values of $\varepsilon_1$, $d_1$, $\varepsilon_{MO}$ and $\alpha$, there generally exist two distinct classes of solutions (and related periodicities) for $d_2$ which ensure total transmission for incidence from a given side; these two solutions merge at a critical value of the ENG-layer electrical thickness, and disappear beyond that point. While the presence of an additional solution does not yield particular advantages, the impossibility of achieving total transmission beyond a critical electrical thickness of the ENG layer does introduce some restrictions in the tuning of the unidirectional resonant tunneling mechanism which, as explained in Sec. \ref{eq:Isolation}, relies on a judicious choice of the ENG parameters.

As another alternative, the tri-layer structure in Fig. \ref{Figure9}(b) can be viewed as an asymmetric variation of that in Fig. \ref{Figure1}, with one of the ENG layers substituted by a DPS layer. This structure is inspired by the ENG-DPS-DPS tri-layers studied in Ref. \onlinecite{Castaldi:2011kg}. Once again, by substituting the central DPS layer with an MO one, the resonant tunneling conditions can be rendered sensitive to the CP handedness. In this case, by straightforward generalization of the results in Ref. \onlinecite{Castaldi:2011kg}, it can be shown that there always exist four independent classes of solutions (and related periodicities) ensuring total transmission for incidence from a given side. Although there is no critical value of the ENG-layer electrical thickness, for increasing opaqueness of the ENG layer, these solutions entail {\em extreme} (either high or low) values of the DPS-material relative permittivity $\varepsilon_3$. While high values of permittivity are technologically-challenging to achieve at optical frequencies, the ENZ behavior is much more manageable, e.g., via the use of recently emerged classes of non-metallic plasmonic materials characterized by low losses and large tunability. \cite{Boltasseva:2011} We verified that isolation performance and structure thickness comparable with those in Figs. \ref{Figure3}--\ref{Figure8} can be obtained.
Therefore, the ENG-MO-DPS configuration in Fig. \ref{Figure9}(b) may constitute an attractive alternative to that in Fig. \ref{Figure1}, by providing additional degrees of freedom in the design.

\section{Conclusions}
\label{Sect:Conclusions}
In this paper, we have studied a unidirectional resonant photon tunneling phenomenon which can occur in tri-layer structures containing ENG and MO material constituents. In particular, we have derived simple analytical conditions and design formulae, which elucidate the underlying physics and allow for an effective parameterization and engineering of the phenomenon. 

Overall, our results indicate that this phenomenon is satisfactorily robust with respect to unavoidable nonidealities, and can be exploited to design compact (wavelength-sized) optical isolators for CP waves. 

Current and future research is aimed at exploring possible practical applications, as well as the study of alternative configurations featuring, e.g., MO metals. \cite{Dong:2010ez,Moccia:2014}


%

\end{document}